\documentclass[12pt]{article}
\usepackage{graphicx}
\usepackage{lscape}
\usepackage{appendix}
\usepackage{multirow}

\begin{document}
\def\qq{\langle \bar q q \rangle}
\def\uu{\langle \bar u u \rangle}
\def\dd{\langle \bar d d \rangle}
\def\sp{\langle \bar s s \rangle}
\def\GG{\langle g_s^2 G^2 \rangle}
\def\Tr{\mbox{Tr}}
\def\figt#1#2#3{
        \begin{figure}
        $\left. \right.$
        \vspace*{-2cm}
        \begin{center}
        \includegraphics[width=10cm]{#1}
        \end{center}
        \vspace*{-0.2cm}
        \caption{#3}
        \label{#2}
        \end{figure}
	}
	
\def\figb#1#2#3{
        \begin{figure}
        $\left. \right.$
        \vspace*{-1cm}
        \begin{center}
        \includegraphics[width=10cm]{#1}
        \end{center}
        \vspace*{-0.2cm}
        \caption{#3}
        \label{#2}
        \end{figure}
                }

\def\ds{\displaystyle}
\def\beq{\begin{equation}}
\def\eeq{\end{equation}}
\def\bea{\begin{eqnarray}}
\def\eea{\end{eqnarray}}
\def\beeq{\begin{eqnarray}}
\def\eeeq{\end{eqnarray}}
\def\ve{\vert}
\def\vel{\left|}
\def\ver{\right|}
\def\nnb{\nonumber}
\def\ga{\left(}
\def\dr{\right)}
\def\aga{\left\{}
\def\adr{\right\}}
\def\lla{\left<}
\def\rra{\right>}
\def\rar{\rightarrow}
\def\lrar{\leftrightarrow}
\def\nnb{\nonumber}
\def\la{\langle}
\def\ra{\rangle}
\def\ba{\begin{array}}
\def\ea{\end{array}}
\def\tr{\mbox{Tr}}
\def\ssp{{\Sigma^{*+}}}
\def\sso{{\Sigma^{*0}}}
\def\ssm{{\Sigma^{*-}}}
\def\xis0{{\Xi^{*0}}}
\def\xism{{\Xi^{*-}}}
\def\qs{\la \bar s s \ra}
\def\qu{\la \bar u u \ra}
\def\qd{\la \bar d d \ra}
\def\qq{\la \bar q q \ra}
\def\gGgG{\la g^2 G^2 \ra}
\def\q{\gamma_5 \not\!q}
\def\x{\gamma_5 \not\!x}
\def\g5{\gamma_5}
\def\sb{S_Q^{cf}}
\def\sd{S_d^{be}}
\def\su{S_u^{ad}}
\def\sbp{{S}_Q^{'cf}}
\def\sdp{{S}_d^{'be}}
\def\sup{{S}_u^{'ad}}
\def\ssp{{S}_s^{'??}}

\def\sig{\sigma_{\mu \nu} \gamma_5 p^\mu q^\nu}
\def\fo{f_0(\frac{s_0}{M^2})}
\def\ffi{f_1(\frac{s_0}{M^2})}
\def\fii{f_2(\frac{s_0}{M^2})}
\def\O{{\cal O}}
\def\sl{{\Sigma^0 \Lambda}}
\def\es{\!\!\! &=& \!\!\!}
\def\ap{\!\!\! &\approx& \!\!\!}
\def\ar{&+& \!\!\!}
\def\ek{&-& \!\!\!}
\def\kek{\!\!\!&-& \!\!\!}
\def\kar{\!\!\!&+& \!\!\!}
\def\cp{&\times& \!\!\!}
\def\se{\!\!\! &\simeq& \!\!\!}
\def\eqv{&\equiv& \!\!\!}
\def\kpm{&\pm& \!\!\!}
\def\kmp{&\mp& \!\!\!}
\def\mcdot{\!\cdot\!}
\def\erar{&\rightarrow&}

\def\vell{ \!\!\! &\ve& \!\!\!}

% .........................................................

\def\simlt{\stackrel{<}{{}_\sim}}
\def\simgt{\stackrel{>}{{}_\sim}}

% .........................................................

\title{
         {\Large
                 {\bf
Tensor form factors of $B \to K_1$ transition 
from QCD light cone sum rules
                 }
         }
      }

\author{\vspace{1cm}\\
{\small T. M. Aliev \thanks {e-mail:
taliev@metu.edu.tr}~\footnote{permanent address:Institute of
Physics,Baku,Azerbaijan}\,\,, M. Savc{\i} \thanks {e-mail:
savci@metu.edu.tr}\,\,, Kwei-Chou Yang \thanks
{e-mail: kcyang@cycu.edu.tw}} \\
{\small Physics Department, Middle East Technical University,
06531 Ankara, Turkey }\\
{\small$^\S$ Department of Physics,Chung Yuan Christian University,} \\
{\small Chung--Li 320, Taiwan}}

\date{}

\begin{titlepage}
\maketitle
\thispagestyle{empty}

\begin{abstract}
The tensor form factors of $B$ into p--wave axial vector meson transition
are calculated within light cone QCD sum rules method. The parametrizations
of the tensor form factors based on the series expansion are presented.
\end{abstract}

%\vspace{1cm}
~~~PACS number(s): 11.55.Hx, 13.75.Gx, 13.75.Jz

\end{titlepage}

\section{Introduction}
Rare decays due to the flavor--changing neutral current $b \to s~(b \to d)$
transitions constitute one of the most important classes of decays
in carefully checking the predictions of the Standard Model (SM) at tree
level, since they are forbidden in SM at loop level. In SM the flavor
changing neutral current (FCNC) processes $b \to s \ell^+ \ell^-$ proceed
through the electroweak penguin and box diagrams. These decays are also very
suitable in looking for new physics (NP) beyond the SM, via contributions of
the new particles to the penguin and box diagrams, that are absent in the SM.

The $B \to K^*(892) \ell^+ \ell^-$ decay has been observed in
\cite{Raasy01,Raasy02}. Moreover, the forward--backward asymmetry has been
measured in \cite{Raasy03,Raasy04}. The longitudinal polarization and
forward--backward asymmetry of  $B \to K^*(892) \ell^+ \ell^-$ and the
isospin asymmetry of $B^0 \to K^{*0}(892) \ell^+ \ell^-$ and $B^\pm \to
K^{*\pm}(892) \ell^+ \ell^-$ are also measured by BaBar Collaboration
in \cite{Raasy05} and \cite{Raasy06}, respectively. 
The experimental results are more or less in
agreement with the predictions of SM. However, the precision of
experiments is currently too low to make the final conclusion. The
situation should considerably be improved at LHCb.

The radiative decays of $B$ meson, involving $K_1(1270)$, where $K_1$ is the
orbitally excited state, is observed by BELLE. The other radiative and
semileptonic decay modes involving $K_1(1270)$ and $K_1(1400)$ are hopefully
expected to be measured soon.

Similar to the  $B \to K^*(892) \ell^+ \ell^-$ decay the $B \to K_1 \ell^+ \ell^-$
decay is also a very good object for probing the new physics effects beyond
the SM. Here the problem becomes more sophisticated due to the mixing of
$K_{1A}~(1^1P_3)$ and $K_{1B}~(1^1P_1)$ state. The physical states
$K_1(1270)$ and $K_1(1400)$ are determined by
\bea
\label{eaasy01}
\left( \begin{array}{l}
\ve K_1(1400) \left.\rra\\
\ve K_1(1270) \left. \rra
\end{array} \right) =
\left( \begin{array}{lr}
\cos\theta & - \sin\theta \\
\sin\theta &   \cos\theta
\end{array} \right)
\left( \begin{array}{l}
\ve K_{1A} \left.\rra\\
\ve K_{1B}\left. \rra
\end{array} \right)~.
\eea
In the present work we calculate the tensor form factors for the $B \to
K_{1A(B)}$ transition in the framework of the light cone QCD sum
rules method (LCSR) (for more about LCSR see \cite{Raasy07,Raasy08}).

The paper is organized in the following way. In section 2 we derive the LCSR
for the tensor form factors  describing the $B \to K_{1A(B)}$
transition. Section 3 is devoted to the numerical analysis of the sum rules for
the form factors. We also summarize our results in this section.

\section{Light cone QCD sum rules for the tensor form factors of the
$B \to K_{1A(B)}$ transition}

The $B \to K_{1A(B)} \ell^+ \ell^-$ decay is described by $b \to s \ell^+
\ell^-$ transition at quark level. The effective Hamiltonian responsible for
the $b \to s \ell^+ \ell^-$ transition is given by,
\bea
\label{eaasy02}
{\cal H} = - 4 {G_F\over \sqrt{2}} V_{tb} V_{ts}^* \sum_{i=1}^{10} C_i(\mu)
{\cal O}(\mu)~,
\eea
where the form of the local Wilson operators ${\cal O}_i~(i=1, \dots, 10)$
is given in \cite{Raasy09}. This effective Hamiltonian leads to the following
result for the $b \to s \ell^+ \ell^-$ decay amplitude
\bea
\label{eaasy03}
{\cal M} \es {G_F\over 2 \sqrt{2}} {\alpha_{\rm em} \over \pi} V_{tb}
V_{ts}^* \Big\{C_9^{\rm eff} \bar{s} \gamma_{\mu}(1-\gamma_5) b \, \bar{\ell}
\gamma_{\mu} \ell + C_{10} \bar{s} \gamma_{\mu}(1-\gamma_5) b \, \bar{\ell}
\gamma_{\mu} \gamma_5 \ell \nnb \\
\ek 2 {m_b \over q^2} C_7 \bar{s} i \sigma_{\mu\nu}
q^\nu (1+ \gamma_5) \bar{\ell} \gamma_{\mu} \ell \Big\}~,
\eea
where the Wilson coefficient $C_9^{\rm eff} = C_9 + Y$, with $Y= Y_{pert}
+ Y_{LD}$, contains both the perturbative and the long
distance contribution parts.
The explicit expression of $C_7$, $C_9$, $Y_{pert}$ and $C_{10}$ are given in 
\cite{Raasy09}. The long distance effects generated by the four--quark
operators with the $c$--quark have recently been calculated for the $B \to
K^\ast \ell^+ \ell^-$ and $B \to K \ell^+ \ell^-$ decays in \cite{Raasy10}
and it is obtained that below the charmonium
region of $q^2$ this effect can change the value of $C_9$ around 5\% and
20\% for $B \to K$ and $B\to K^*$ transitions, respectively. Similar
calculations for $B\to K_1$ transition has not yet been calculated. For
simplicity, in the following discussions we denote $K_{1A}$ and $K_{1B}$
as $K_1$. 

It follows from Eq.(\ref{eaasy03}) that for the calculation of the $B \to
K_1$ transition, the matrix elements $\lla
K_1(p,\lambda) \vel \bar{s} \gamma_\mu (1-\gamma_5)b \ver B(p_B)\rra$ and 
$\lla K_1(p,\lambda) \vel \bar{s} \sigma_{\mu\nu} q^\nu (1+\gamma_5)b \ver
B(p_B) \rra$ are needed. For the $B \to K_1$ transition, 
these matrix elements are defined in terms of the
form factors as follows:
\bea
\label{eaasy05}
\lla K_1(p,\lambda) \vel \bar{s} \gamma_\mu (1-\gamma_5) b \ver B(p_B) \rra
\es - i {2 \over m_B+m_{K_1}} \epsilon_{\mu\nu\alpha\beta}
\varepsilon^{(\lambda)*\nu} p_B^\alpha p^\beta A^{K_1} (q^2) \nnb \\
\ek \Bigg[ (m_B + m_{K_1}) \varepsilon_\mu^{(\lambda)*} V_1^{K_1}(q^2) -
(p_B+p)_\mu (\varepsilon^{(\lambda)*} p_B) {V_2^{K_1}(q^2) \over m_B +
m_{K_1}} \Bigg] \nnb \\
\ar 2 m_{K_1} {( \varepsilon^{(\lambda)*}p_B) \over q^2} q_\mu
[V_3^{K_1}(q^2) - V_0^{K_1}(q^2)]~, \\ \nnb \\
\label{eaasy06}
\lla K_1(p,\lambda) \vel \bar{s} \sigma_{\mu\nu} q^\nu(1+\gamma_5) b \ver B(p_B) \rra
\es 2 T_1^{K_1} (q^2) \epsilon_{\mu\nu\alpha\beta}
\varepsilon^{(\lambda)*\nu} p_B^\alpha p^\beta \nnb \\
\ek i T_2^{K_1} (q^2) \Big[(m_B^2 -
m_{K_1}^2) \varepsilon^{(\lambda)*}_\mu - (\varepsilon^{(\lambda)*} q)
(p_B+p)_\mu \Big] \nnb \\
\ek  i T_3^{K_1} (q^2) (\varepsilon^{(\lambda)*} q)\Bigg[ q_\mu
- {q^2 \over m_B^2 - m_{K_1}^2} (p_B+p)_\mu \Bigg],
\eea
where $q = p_B-p$.
There are the following relations between the form factors:
\bea
\label{eaasy07}
V_3^{K_1}(q^2) \es {m_B + m_{K_1} \over 2 m_{K_1}} V^{K_1}_1(q^2) - {m_B - m_{K_1} \over 2 m_{K_1}}
V_2^{K_1}(q^2) ~, \nnb \\
V_3^{K_1}(0) \es V^{K_1}_0(0),~~ \mbox{\rm and},\nnb \\
T_1^{K_1}(0) \es T^{K_1}_2(0)~.
\eea

To be able to calculate the form factors responsible for the  $B \to
K_1$ transition we consider the following two correlation functions:
\bea
\label{eaasy08}
\Pi_\mu \es i \int d^4x e^{iqx} \lla K_1 (p, \lambda) \vel
T\{\bar{s}(x) \gamma_\mu (1-\gamma_5) b(x)\, \bar{b}(0) i \gamma_5 d(0) \}\ver 0 \rra~, \\
\label{eaasy09}
\Pi_{\mu\nu} \es i \int d^4x e^{iqx} \lla K_1 (p, \lambda) \vel
T\{\bar{s}(x) \sigma_{\mu\nu} \bar{b}(x)\, b(0) i \gamma_5 d(0) \}\ver 0 \rra~.
\eea
In order to construct the sum rules for the form factors responsible for
the $B \to K_1$ transition these correlation functions should be
calculated in two different languages, in terms of hadrons and quark and
gluon degrees of freedom. The calculation of the correlation function in
terms of quark and gluon degrees of freedom is carried out at virtualities 
$m_b^2-p_B^2 \ge \Lambda_{QCD} m_b$ and $m_b^2-q^2 \ge \Lambda_{QCD} m_b$. 
Using the operator product expansion, the sum
rules are obtained by equating these two representations through the
dispersion relations.

Phenomenological parts of the correlation functions (\ref{eaasy08}) and
(\ref{eaasy09}) can be obtained by inserting complete set of hadrons with
the same quantum numbers as the interpolating current, and separating the
ground state one can easily obtain
\bea
\label{eaasy10}
\Pi_\mu \es -{\lla K_1 (p,\lambda) \vel \bar{s}
\gamma_\mu (1-\gamma_5) b \ver B(p_B) \rra \lla B(p_B) \vel
\bar{b} i \gamma_5 d \ver 0 \rra \over p_B^2 - m_B^2} + \cdots~, \\
\label{eaasy11}
\Pi_{\mu\nu} \es - {\lla K_1 (p,\lambda) \vel \bar{s}
\sigma_{\mu\nu} b \ver B(p_B) \rra \lla B(p_B) \vel
\bar{b} i \gamma_5 d \ver 0 \rra \over p_B^2 - m_B^2} + \cdots~,
\eea
where ``$\cdots$" describes the contributions coming from higher states and
continuum, and the matrix element $\lla K_1 (p,\lambda) \vel \bar{s}
\gamma_\mu (1-\gamma_5) b \ver B(p_B) \rra$ is given in Eq. (\ref{eaasy05}).
The second matrix element in  Eq. (\ref{eaasy10}) is expressed in the
standard way
\bea
\label{eaasy12}
\lla B(p_B) \vel \bar{b} i \gamma_5 d \ver 0 \rra = {f_B m_B^2 \over m_b}~,
\eea
where $f_B$ is the $B$--decay constant and $m_b$ is the $b$--quark mass. 
The matrix element $\lla K_1 (p,\lambda) \vel \bar{s} \sigma_{\mu\nu} b 
\ver B(p_B) \rra$ is defined as
\bea
\label{eaasy13}
\lla K_1 (p,\lambda) \vel \bar{s} \sigma_{\mu\nu} b \ver B(p_B) \rra \es
-i A(q^2) [\varepsilon_\mu^{(\lambda)*} (p+p_B)_\nu - \varepsilon_\nu^{(\lambda)*}
(p+p_B)_\mu ] \nnb \\
\ar i B(q^2) (\varepsilon_\mu^{(\lambda)*} q_\nu - \varepsilon_\nu^{(\lambda)*} q_\mu) 
 + i \frac{2C(q^2)}{m_B^2-m_{K_{1}}} (p_\mu q_\nu - p_\nu q_\mu)~ . ~~
\eea
Contracting Eq. (\ref{eaasy13}) with the momentum $q^\nu$ and using the
relation
\bea
\label{nolabel}
\sigma_{\mu\nu} \gamma_5 = - {i\over 2} \epsilon_{\mu\nu\alpha\beta}
\sigma^{\alpha\beta}~,\nnb
\eea
the following relations among $A,~B$ and $C$ can easily be obtained:
\bea
\label{eaasy14}
T_1^{K_1}(q^2) \es A(q^2)~, \nnb \\
T_2^{K_1}(q^2) \es A(q^2)- {q^2\over m_B^2-m_{K_1}^2} B(q^2)~, \nnb \\
T_3^{K_1}(q^2) \es B(q^2) + C(q^2)~.
\eea
Using Eqs. (\ref{eaasy12}) and (\ref{eaasy13}), for the phenomenological
parts of the correlation functions we get
\bea
\label{eaasy15}
\Pi_\mu \es -{f_B m_B^2 \over m_b} {1\over p_B^2 - m_B^2}
\Bigg\{ - {2i \over  m_B^2 -
m_{K_1}^2} \epsilon_{\mu\nu\alpha\beta} \varepsilon^{(\lambda)*\nu} p_B^\alpha
p^\beta A^{K_1}(q^2) \nnb \\
\ek \Bigg[(m_B+m_{K_1}) \varepsilon_\mu^{(\lambda)*} V_1^{K_1}(q^2) - P_\mu (\varepsilon^{(\lambda)*} q)
{V_2^{K_1}(q^2)\over m_B+m_{K_1}} \Bigg]  \nnb \\
 \ar 2 m_{K_1} {\varepsilon^{(\lambda)*} q \over q^2} q_\mu
[V_3^{K_1}(q^2)-V_0(q^2)] \Bigg\}~, \\ \nnb \\
\label{eaasy16}
\Pi_{\mu\nu} \es - {f_B m_B^2 \over m_b} {1\over p_B^2 -
m_B^2} \Bigg\{ - i
A(q^2) (\varepsilon_\mu^{(\lambda)*} P_\nu - \varepsilon_\nu^{(\lambda)*} P_\mu) +
i B(q^2) (\varepsilon_\mu^{(\lambda)*} q_\nu - \varepsilon_\nu^{(\lambda)*} q_\mu) \nnb \\
\ar 2i C(q^2) {\varepsilon^{(\lambda)*} q \over  m_B^2-m_{K_1}^2} (p_\mu q_\nu - p_\nu
q_\mu)\Bigg\}~,
\eea
where $P=p_B + p$.

We now proceed to calculate the theoretical part of the correlation
functions. The calculation is performed by using the background field
approach \cite{Raasy11}. In the large virtuality region, where $m_b^2-p_B^2
 \gg \Lambda_{QCD} m_b$ and $m_c^2-q^2 \gg \Lambda_{QCD} m_b$, 
the operator product expansion is applicable to the correlation functions.
In light cone sum rules the method is based on the expansion of the non--local
quark--antiquark operators in powers of the deviation from the light cone. In
obtaining  the expression of the correlation functions the propagator of
heavy quark and the matrix elements of the non--local operators
$\bar{q}(x_1) \Gamma_i q(x_2)$ and $\bar{q}(x_1) G_{\mu\nu} q(x_2)$ between
the vacuum and axial--vector meson are needed, where $\Gamma_i$ are the Dirac
matrices (in our case $\gamma_\mu (1-\gamma_5)$ or $\sigma_{\mu\nu}$), and
$G_{\mu\nu}$ is the gluon field strength tensor.

The expression of the heavy quark operator is given in \cite{Raasy12}
\bea
\label{eaasy17}
S_Q = S_Q^{free} (x)  + ig_s \int {d^4k\over (2 \pi)^4} e^{-ikx} \int du
\Bigg[{\rlap/{k} + m_b\over 2 (m_b^2-k^2)^2} G_{\mu\nu}(ux) \sigma^{\mu\nu} +
{u\over m_b^2-k^2} x_\mu G^{\mu\nu} (ux) \gamma_\nu \Bigg]~,
\eea
where $S_Q^{free}$ is the free quark operator and we adopt the convention for covariant derivative
$D_\alpha=\partial_\alpha +ig_s A^a_\alpha \lambda^a/2$.

Two particle distribution amplitudes for the axial vector mesons
are presented in  \cite{Raasy13,Raasy14}
\bea
\label{eaasy18}
\lla K_1(p,\lambda) \vel \bar{s}_\alpha (x) q_\delta (0) \ver 0 \rra \es
-{i\over 4} \int du e^{iupx} \nonumber\\
 & \times & \Bigg\{ f_{K_1} m_{K_1} \Bigg[ \rlap/{p}\gamma_5
{\varepsilon^{(\lambda)*} x\over px} \phi_\parallel (u) + \Bigg(\rlap/{\varepsilon}^{(\lambda)*} -
{\varepsilon^{(\lambda)*} x \over px} \rlap/{p} \Bigg) \gamma_5 g_\perp^{(a)}(u) \nnb \\
\ek \rlap/{x} \gamma_5 {\varepsilon^{(\lambda)*} x \over 2(px)^2} m_{K_1}^2 \bar{g}_3(u) +
\epsilon_{\mu\nu\rho\sigma} \varepsilon^{(\lambda)*\nu} p^\rho x^\sigma \gamma^\mu
{g_\perp^{(v)} \over 4} \Bigg] \nnb \\
\ar f_\perp^A \Bigg[ {1\over 2} (\rlap/{p} \rlap/{\varepsilon}^{(\lambda)*} -
\rlap/{\varepsilon}^{(\lambda)*} \rlap/{p} ) \gamma_5 \phi_\perp (u) - {1\over 2}
(\rlap/{p} \rlap/{x} - \rlap/{x} \rlap/{p}) \gamma_5 {\varepsilon^{(\lambda)*} x \over
(px)^2} m_{K_1}^2 \bar{h}_\parallel^{(t)} (u) \nnb \\
\ek {1\over 4}
(\rlap/{\varepsilon}^{(\lambda)*} \rlap/{x} - \rlap/{x} \rlap/{\varepsilon}^{(\lambda)*})
\gamma_5 {m_{K_1}^2 \over px} \bar{h}_3 (u) + i (\varepsilon^{(\lambda)*} x) m_{K_1}^2 \gamma_5
{h_\parallel^{(p)} (u) \over 2} \Bigg] \nonumber\\ 
&+& {\cal O}(x^2)\Bigg\}_{\delta\alpha}~, 
\eea
where
\bea
\label{eaasy1818}
\bar{g}_3 (u) \es g_3(u) + \phi_\parallel - 2 g_\perp^{(a)} (u) ~, \nnb \\
\bar{h}_\parallel^{(t)}(u) \es h_\parallel^{(t)}(u) - {1\over 2} \phi_\perp (u)
- {1\over 2} h_3 (u)~, \nnb \\
\bar{h}_3 (u) \es h_3 (u) - \phi_\perp (u)~,
\eea
and $\phi_\parallel$, $\phi_\perp$ are the twist--2, $g_\perp^{(a)}$,
$g_\perp^{(v)}$, $h_\parallel^{(t)}$ and $h_\parallel^{(p)}$ are twist--3, and
$g_3$ and $h_3$ are twist--4 functions. The three particle distribution
amplitudes are define as
\bea
\label{eaasy19}
\lla K_1(p,\lambda) \vel \bar{s}(x) \gamma_\alpha \gamma_5 g_s
G_{\mu\nu} (ux) q (0) \ver 0 \rra \es p_\alpha (p_\nu
\varepsilon^{(\lambda)*}_\mu - p_\mu \varepsilon^{(\lambda)*}_\nu) f_{3 K_1}^A {\cal A} +
\cdots~,\nnb \\
\lla K_1(p,\lambda) \vel \bar{s}(x) \gamma_\alpha g_s
\widetilde{G}_{\mu\nu} (ux) q (0) \ver 0 \rra \es i p_\alpha (p_\mu
\varepsilon^{(\lambda)*}_\nu - p_\nu \varepsilon^{(\lambda)*}_\mu) f_{3 K_1}^V {\cal V} +
\cdots~,
\eea
where
\bea
\label{nolabel}
{\cal A} = \int {\cal D}\alpha e^{iPx (\alpha_1 + u \alpha_3)} {\cal
A}(\alpha_1,\alpha_2,\alpha_3)~,\nnb
\eea
and
\bea
\label{nolabel}
\int {\cal D}\alpha = \int_0^1 d \alpha_1 \int_0^1 d \alpha_2 \int_0^1 d
\alpha_3~ \delta(1-\alpha_1-\alpha_2-\alpha_3)~. \nnb
\eea
Here  $\alpha_1$, $\alpha_2$, and $\alpha_3$ are the
respective momentum fractions carried by $s$, $\bar{q}$ quarks
and gluon in the meson. Using these definitions, and after lengthy calculations for the theoretical
parts of the correlation functions, we obtain
\bea
\label{eaasy20}
&&Correlation~function = \nnb \\
&&{1\over 4} \int du \Big\{ f_{K_i} m_{K_i} \Big[ \varepsilon_\alpha^{(\lambda)*}
\Phi_a^{(i)} {\partial\over \partial Q_\alpha} \mbox{Tr}(\Gamma \rlap/{P}
S_Q ) - g_\perp^a \mbox{Tr}(\Gamma S_Q \rlap/{\varepsilon}^{(\lambda)*}) \nnb \\
&&- {1\over 2} m_{K_i}^2 \bar{g}_3^{(ii)} \varepsilon_\alpha^{(\lambda)*} {\partial\over
\partial Q_\alpha} {\partial\over \partial Q_\beta} \mbox{Tr}(\Gamma
S_Q \gamma_\beta)
+ i \epsilon_{\alpha\beta\rho\sigma} {g_\perp^v \over 4}
\varepsilon_\beta^{(\lambda)*} p^\rho  {\partial\over \partial Q_\sigma}
\mbox{Tr}(\Gamma S_Q \gamma_\alpha \gamma_5) \Big] \nnb \\
&& + f^\perp_{K_i} \Big[ {1\over 2} \phi_\perp (u) \mbox{Tr}[\Gamma S_Q
(\rlap/{p} \rlap/{\varepsilon}^{(\lambda)*}  - \rlap/{\varepsilon}^{(\lambda)*} \rlap/{p})] -
{1\over 2} m_{K_i}^2 \bar{h}_\parallel^{(t)(ii)} \varepsilon_\alpha^{(\lambda)*}
 {\partial\over \partial Q_\alpha}  {\partial\over \partial Q_\beta}
\mbox{Tr}[\Gamma S_Q (\rlap/{p} \gamma_\beta - \gamma_\beta \rlap/{p})] \nnb \\
&& + {h_3^{(i)}\over 4} m_{K_i}^2  {\partial\over \partial Q_\alpha} \mbox{Tr}[\Gamma
S_Q \gamma_5 (\rlap/{\varepsilon^{(\lambda)*}} \gamma_\alpha - \gamma_\alpha
\rlap/{\varepsilon}^{(\lambda)*})] + {1\over 2} h_\parallel^{(p)} m_{K_i}^2
\varepsilon^{(\lambda)*}_\alpha {\partial\over \partial Q_\alpha}  \mbox{Tr}[\Gamma
S_Q \Big] \Big\} \nnb \\
&& + {1\over 4} \int dv \int{\cal D}\alpha_i {1\over \{m_b^2 - [q + (\alpha_1 +
v \alpha_3) p]^2 \}^2} \Big\{ 2 v pq \Big[f_{3 i}^A {\cal A}(\alpha_i) + f_{3 i}^V
{\cal V}(\alpha_i) \Big] \mbox{Tr}(\Gamma \rlap/{\varepsilon}^{(\lambda)*} \rlap/{p}) \Big\}~,\nonumber\\
\eea
where
\bea
\label{nolabel}
S_Q \es {m_b + \rlap/{Q} \over m_b^2-Q^2}~,~~{\mbox \rm with}~,~~~
Q=q+pu~,\nnb \\
\Phi_a^{(i)} \es \int_0^u \Big[ \phi_\parallel^{(v)} - g_\perp^a (v) \Big]
dv~, \nnb \\
f^{(i)} \es \int_0^u f(v) dv~, \nnb \\
f^{(ii)} \es \int_0^u dv \int_0^v dv^\prime f(v^\prime)~, \nnb
\eea
and $i=1~(2)$ correspond to $K_{1A}~(K_{1B})$, respectively,
$\Gamma$ is equal to $\gamma_\mu (1-\gamma_5)$ or $\sigma_{\mu\nu}$.
After taking derivatives and traces, equating expressions of correlation
functions (\ref{eaasy15}), (\ref{eaasy16}) and (\ref{eaasy20}), and performing
Borel transformation with respect to the variable $-(p+q)^2$ in order to
suppress the higher states and continuum contributions, one can obtain the
sum rules for the transition form factors.
Here we present the sum rules only for the tensor form factors, since 
$V_1^{K_1}$, $V_2^{K_1}$, $V_0^{K_1}$ and $A^{K_1}$ are calculated within the same framework in
\cite{Raasy15}:
\bea
\label{eaasy21}
A_i(q^2) \es - {m_b^2 f_{\perp i} \over 2 m_B^2 f_B}
e^{m_B^2/M^2} \int_0^1 du \Bigg\{{1\over u} e^{-s(u)/M^2}
\theta[s_0-s(u)] \Bigg[ \phi_\perp (u) \nnb \\
\ek  {m_i f_i\over m_b f_{\perp i}}
\Bigg( u g_\perp^{(a)} (u) + \phi_a^{(i)} + {g_\perp^v (u) \over 4} \Bigg)
\Bigg] - {1\over 4} {e^{-s(u)/M^2} m_i f_i \over u m_b f_{\perp i}} (m_b^2 +
q^2) g_\perp^v (u) \nnb \\
\cp \Bigg( {\theta[s_0-s(u)] \over u M^2} + {\delta[u-u_0] \over s_0-q^2}\Bigg) \Bigg\}
\nnb \\
\ek {m_b \over 2 m_B^2 f_B} e^{m_B^2/M^2}
\int_0^1 v dv \int e^{-s(k)/M^2} d\alpha_1 d\alpha_3 {f_{3i}^A {\cal A}(\alpha_i)
+ f_{3i}^V {\cal V}(\alpha_i) \over (\alpha_1+ v \alpha_3)^2} \nnb \\
\cp \Bigg\{
\theta[s_0-s(k)] - (m_b^2 - q^2) \Bigg( {\theta[s_0-s(k)]  \over (\alpha_1+
v \alpha_3) M^2} + {\delta [k-u_0] \over s_0-q^2}\Bigg) \Bigg\}~, \nnb \\ \nnb \\
B_i(q^2) \es  - { m_b^2 f_{\perp i} \over 2 m_B^2
f_B} e^{m_B^2/M^2} \int_0^1 du \Bigg\{ {1\over u}
e^{-s(u)/M^2} \theta[s_0-s(u)] \Bigg[ \phi_\perp (u) \nnb \\
\ek  {m_i f_i\over m_b f_{\perp i}} \Bigg(-(2-u) g_\perp^{(a)} (u) +
\phi_a^{(i)} + \Bigg(1-{2\over u} \Bigg) {g_\perp^v (u) \over 4} \Bigg)
\Bigg] \nnb \\
\ek {1\over u} e^{-s(u)/M^2} {1\over 4} {m_i f_i \over m_b f_{\perp i}} \Bigg[
2 m_b^2 - (m_b^2-q^2) \Bigg(1-{2\over u} \Bigg) \Bigg] g_\perp^{(v)} (u)
\nnb \\
\cp \Bigg( {\theta[s_0-s(u)] \over u M^2} + {\delta[u-u_0] \over s_0-q^2}\Bigg) \Bigg\}
\nnb \\
\ek {m_b \over 2 m_B^2 f_B} e^{m_B^2/M^2}
\int_0^1 v dv \int e^{-s(k)/M^2} d\alpha_1 d\alpha_3 {f_{3i}^A {\cal A}(\alpha_i)
+ f_{3i}^V {\cal V}(\alpha_i) \over (\alpha_1+ v \alpha_3)^2} \nnb \\
\cp \Bigg\{
\theta[s_0-s(k)] - (m_b^2 - q^2) \Bigg( {\theta[s_0-s(k)]  \over (\alpha_1+
v \alpha_3) M^2} + {\delta [k-u_0] \over s_0-q^2} \Bigg) \Bigg\}~, \nnb \\ \nnb \\
C_i(q^2) \es {m_b m_i f_i \over 2 f_B} e^{m_B^2/M^2}
\int_0^1 du \Bigg\{{1\over u} e^{-s(u)/M^2}
 \Bigg[ \Bigg( 2 \phi_a^{(i)} (u) - {g_\perp^{(v)} (u) \over 2} \Bigg) \nnb \\
\cp \Bigg( {\theta[s_0-s(u)] \over u M^2} + {\delta[u-u_0] \over s_0-q^2}\Bigg)\Bigg\}~,
\eea
where $i=A$ or $B$, and
\bea
\label{nolabel}
s(n) = {m_b^2 - (1-n) q^2 \over n}\,, \qquad k\equiv \alpha_1 +v \alpha_3
\,, \mbox{\rm ~and}\,,~ \qquad u_0 = {m_b^2-q^2 \over s_0-q^2}~. \nnb
\eea
In these expressions, we neglect terms $\sim m_{K_i}^2$. Using Eq.
(\ref{eaasy14}) one can easily obtain the corresponding sum rules for $T_1$,
$T_2$ and $T_3$ tensor form factors.

%
%  ---------------- april 11, 2011 ------------------
%
A few words about the form factors responsible for the $B \to K_1$
transition, in the large recoil region in the heavy quark limit, are in order.
It can be shown that, similar to the $B \to V$ (vector meson) case,
all seven form factors responsible for the $B \to K_1$ transition 
can be expressed in terms of the independent functions $\xi_\perp^{K_1}(q^2)$ and 
$\xi_\parallel^{K_1}(q^2)$, in
the large recoil region and in the heavy quark limit. Indeed we find that,
for to the $B \to K_1$ transition these form factors can be written in terms
of $\xi_\perp^{K_1}(q^2)$ and 
$\xi_\parallel^{K_1}(q^2)$ as:
\bea
\label{eaasy22}
V_0^{K_1}(q^2) \es \Bigg(1 - {m_{K_1}^2 \over m_B E}\Bigg)
\xi_\parallel^{K_1}(q^2) + {m_{K_1} \over m_B} \xi_\perp^{K_1}(q^2)~,\nnb \\
V_1^{K_1}(q^2) \es \Bigg( {2 E \over m_B + m_{K_1}} \Bigg)
\xi_\perp^{K_1}(q^2)~,\nnb \\
V_2^{K_1}(q^2) \es \Bigg(1 + {m_{K_1} \over m_B}\Bigg)
\xi_\perp^{K_1}(q^2)- {m_{K_1} \over E} \xi_\parallel^{K_1}(q^2)~,\nnb \\
A^{K_1}(q^2) \es \Bigg(1 + {m_{K_1} \over m_B}\Bigg)
\xi_\perp^{K_1}(q^2)~,\nnb \\
T_1^{K_1}(q^2) \es \xi_\perp^{K_1}(q^2)~,\nnb \\
T_2^{K_1}(q^2) \es \Bigg(1 - {q^2 \over m_B^2 - m_{K_1}^2}\Bigg)
\xi_\perp^{K_1}(q^2)~,\nnb \\
T_3^{K_1}(q^2) \es \xi_\perp^{K_1}(q^2) - \Bigg(1 - 
{m_{K_1}^2 \over m_B^2}\Bigg) {m_{K_1} \over E} \xi_\parallel^{K_1}~,
\eea
where
\bea
\label{nolabel}
E = {m_B^2 + m_{K_1}^2 - q^2\over 2 m_B}~, \nnb
\eea
is the energy of $K_1$ meson.

Explicit expressions of the functions $\xi_\perp^{K_1}(q^2)$ and
$\xi_\parallel^{K_1}(q^2)$ can be obtained following the same steps of
calculation as is given in \cite{Raasy16}. These expressions are quite
lengthy, and therefore we do not present them here in this work.

Our final remark in this section is as follows. The physical $K_1(1270)$ and
$K_1(1400)$ are the mixing states of $K_{1A}$ and $K_{1B}$, and the form
factors for the $B \to K_1(1270)$ and $B \to K_1(1400)$ transitions can be
obtained from $B \to K_{1A}$ and $B \to K_{1B}$ transition form factors with
the help of following transformations,
\bea
\label{nolabel}
\left( \begin{array}{c}
\lla \bar{K}_1(1270) \vel \bar{s} \gamma_\mu (1-\gamma_5) b \ver B \rra \\
\lla \bar{K}_1(1400) \vel \bar{s} \gamma_\mu (1-\gamma_5) b \ver B \rra
\end{array} \right) =
{\cal M}
\left( \begin{array}{c}
\lla \bar{K}_{1A} \vel \bar{s} \gamma_\mu (1-\gamma_5) b \ver B \rra \\
\lla \bar{K}_{1B} \vel \bar{s} \gamma_\mu (1-\gamma_5) b \ver B \rra
\end{array} \right)~, \nnb
\eea

\bea
\label{nolabel}
\left( \begin{array}{c}
\lla \bar{K}_1(1270) \vel \bar{s} \sigma_{\mu\nu} q^\nu (1+\gamma_5) b \ver B \rra \\
\lla \bar{K}_1(1400) \vel \bar{s} \sigma_{\mu\nu} q^\nu (1+\gamma_5) b \ver B \rra
\end{array} \right) =
{\cal M}
\left( \begin{array}{c}
\lla \bar{K}_{1A} \vel \bar{s} \sigma_{\mu\nu} q^\nu (1+\gamma_5) b \ver B \rra \\
\lla \bar{K}_{1B} \vel \bar{s} \sigma_{\mu\nu} q^\nu (1+\gamma_5) b \ver B \rra
\end{array} \right)~, \nnb
\eea
where
\bea
{\cal M} =
\left( \begin{array}{lr}
\sin\theta &  \cos\theta \\
\cos\theta   & -\sin\theta
\end{array} \right)~. \nnb
\eea

From the analysis of $B \to K_1 \gamma$ and $\tau^- \to K_1(1270) \nu_\tau$
decays, the mixing angle $\theta$ is obtained to have the value $\theta
= -(34^0 \pm 13^0)$, where the minus sign is related to the relative phases
of $\ve \bar{K}_{1A} \left.\rra$ and $\ve \bar{K}_{1B} \left.\rra$. The
phases are fixed by adopting the conventions $f_{K_{1A}} > 0$ and
$f_{K_{1B}} > 0$.

\section{Numerical analysis}

In this section we present our numerical analysis of the sum rules for the
form factors. The main parameters entering to the sum rules for the $B\rar
K_1$ transition form factors are the leptonic decay constant of the $B_d$
meson, DAs of the $K_1$ meson, the mass of the b--quark, Borel parameter
$M^2$ and and the continuum threshold $s_0$.
The explicit expressions of the DAs for $K_{1A}$ and $K_{1B}$ mesons, as
well as the parameters in the DAs are given in \cite{Raasy13,Raasy14} and
their properties are presented them in the Appendix.
%
%  ---------------- march 11, 2011 ------------------
%

Few words about the value of the leptonic decay constant $f_B$ are in order.
It is shown in \cite{Raasy17} that the pole mass of the $b$--quark produces
rather large higher--order radiative NLO corrections in the results for
$f_B$. Moreover, it is shown in this work that in the $\overline{MS}$ scheme the
higher order corrections are under control, and therefore, the predictions
for $f_B$ is more reliable. For this reason in further numerical analysis we
use the value $f_B = 210 \pm 19~MeV$ as is obtained in \cite{Raasy17} within
the framework of $\overline{MS}$ scheme, at $\mu=\overline{m}_b$ scale. For
the $\overline{MS}$ mass $\overline{m}_b$ we use $\overline{m}_b(2~GeV) =
4.98~GeV$ which is obtained from $\overline{m}_b(\overline{m}_b) = 4.2 ~GeV$
\cite{Raasy17}. For the strange quark mass we use the value
$m_s(2~GeV)=102 \pm 8~MeV$, which is obtained from the analysis of
pseudoscalar QCD sum rules \cite{Raasy18}.

%
% -----------------------------------------------------
%                      
As has already been
noted, the sum rules for the
form factors also contain two auxiliary parameters: the Borel mass parameter
$M^2$ and the continuum threshold $s_0$, and obviously, any physical
quantity should be independent of them. For this reason, we try to find such
regions these parameters where physically measurable quantities are
independent of them. 

In determining the value of the continuum threshold $s_0$, we require that
the prediction of the mass sum rules for the $B$ meson coincides with the
experimental data. We also require that $s_0$ must
not be far from the ``reliable" region, i.e., it should be below the next
resonance in this channel. These conditions lead to the result that, the
expected values of $s_0$ lie in the interval $33~GeV^2 \le s_0 \le
38~GeV^2$. The upper bound of $M^2$ is determined by demanding that the
total contributions of higher states and continuum threshold should be less
than half of the dispersion integral. The lower limit of $M^2$ is found if
we require that the highest power $1/M^2$ term contributes less than, say,
25\% of the sum rules. Both these conditions are satisfied when $M^2$ varies
in the region $6~GeV^2 \le M^2 \le 16~GeV^2$. In numerical calculations we
have used $M^2=10~GeV^2$ and $s_0=34~GeV^2$. 

Unfortunately, the sum rules cannot predict the dependence of the form
factors on $q^2$ in the relevant physical region $4 m_\ell^2 \le q^2 \le
(m_B-m_{K_1})^2$. The sum rules results are not reliable when $q^2 >
10~GeV^2$. 
In order to extend the results for
the form factors coming from QCD sum rules predictions to cover the whole
physical region, we look for a parametrization of the form factors in such a
way that, the results obtained for the region $4 m_\ell^2 \le q^2 \le
10~GeV^2$ can be extrapolated to whole physical region.

%
%  ---------------- march 21, 2011 ------------------
%
To extend our calculations to whole physical region, we use the $z$--series
parametrization (for more about this parametrization, see \cite{Raasy19} and
references therein), 
which is based on the
analyticity of the form factors on $q^2$. Before presenting the series
expansion of the tensor form factors of the $B \to K_1$ transition,
following the work \cite{Raasy19}, we present the helicity amplitudes as the 
linear combination of the form factors as are defined in Eq. (\ref{eaasy05}),
which are more convenient for the analysis. After a simple calculation we
obtain the following helicity amplitudes for the $B \to K_1$ transition (see
also \cite{Raasy19})
\bea
\label{eassy23}
H_0 (q^2) \es {\sqrt{q^2} (m_B^2 + 3 m_{K_1}^2 - q^2) \over 2 m_{K_1}
\sqrt{\lambda}} T_2(q^2) - {\sqrt{q^2 \lambda} \over 2 m_{K_1} (m_B^2 -
m_{K_1}^2)} T_3 (q^2)~, \nnb \\
H_1 (q^2) \es \sqrt{2} T_1(q^2)~, \nnb \\
H_2(q^2) \es {\sqrt{2} \over \sqrt{\lambda}} (m_B^2 - m_{K_1}^2) T_2
(q^2)~,
\eea
where $\lambda(m_B^2,m_{K_1}^2,q^2)=m_B^4+m_{K_1}^4 +q^4-2 m_B^2 m_{K_!}^2
-2 m_B^2 q^2 -2 m_{K_1}^2 q^2$, and subscripts $0$ and $1,~2$ correspond to 
the longitudinal and linear
combinations of the transversal polarizations,
\bea
\label{nolabel}
\varepsilon_{1,2}^\mu = {1\over \sqrt{2}} \Big[ \varepsilon_-^\mu(q) \mp
\varepsilon_+^\mu(q) \Big]~, \nnb
\eea
of the virtual axial meson, respectively.  

We define the following parametrization for the tensor form factors based on
the z--series expansion,
\bea
\label{eaasy24}
H_0(q^2) \es {\sqrt{-z(q^2,0)} \over B(q^2) \sqrt{z(q^2,q_-^2)} \phi(q^2)}
\sum_{k=0}^{N-1} \beta_k^{(0)} z^k~, \nnb \\
H_1(q^2) \es {1\over B(q^2) \phi(q^2)} \sum_{k=0}^{N-1} \beta_k^{(1)} z^k~,
\nnb \\
H_2(q^2) \es {1 \over B(q^2) \sqrt{z(q^2,q_-^2)} \phi(q^2)}
\sum_{k=0}^{N-1} \beta_k^{(2)} z^k~,
\eea
where $z=z(q^2,q_0^2)$ is defined as,
\bea
\label{eaasy25}
z(q^2,q_0^2) = {\sqrt{q_+^2 - q^2} - \sqrt{q_+^2 - q_0^2} \over
\sqrt{q_+^2 - q^2} + \sqrt{q_+^2 - q_0^2} }~,
\eea
and $q_\pm^2 = (m_B \pm m_{K_1})^2$ and $B(q^2) = z(q^2,m_{res}^2$). 
The parameter $q_0^2$ is chosen from the solution of the equation 
$z(0,q_0^2) = - z(q_-^2,q_0^2)$, which gives $q_0^2=10.55~GeV^2$.
The mass of the resonances entering to the factor $B(q^2)$ are $m_B^* (1^-)=
5.41~GeV$ for $H_0(q^2),~H_2(q^2)$, and $m_B^* (1^-)= 5.83~GeV$ 
for $H_1(q^2)$, respectively (see \cite{Raasy19}). The function $\phi(q^2)$
is given in Eq. (39) of the work \cite{Raasy19}. In order to perform the
numerical analysis for the helicity amplitudes, and derive the unitary
bound one needs to calculate the two--point correlation function of the
tensor current. This calculation is done in \cite{Raasy19} and we will use
the results of this work. For the $z$--series expansion parametrization, the
unitarity constraint leads to the result
\bea
\label{eaasy26}
\sum_{k=0}^{N-1} \Big\{ \beta_k^{(0)2} + \beta_k^{(1)2} + \beta_k^{(2)2}
\Big\} \le 1~.
\eea

In Figs. (1)--(3) we present the fits of the series expansion
parametrization to LCSR results for the helicity amplitudes $H_0(q^2)$,
$H_1(q^2)$ and $H_2(q^2)$, respectively. In these figures, we take into
account the uncertainties coming from Gegenbauer moments of the axial vector
meson, mass of $b$ and $s$ quark, Borel mass $M^2$ and the threshold
parameter $s_0$. 

In Table 1 we present the values of the coefficients $\beta_k^{(0)K_{1A}}$
and $\beta_k^{(1)K_{1A}}$, $(k=0,1)$, entering to the 
series expansion of the helicity amplitudes $H_0(q^2)$, $H_1(q^2)$ and 
$H_2(q^2)$, and the unitarity constraint $\sum_{k=0}^1{(\beta_k^{K_{1A}})^2}$
given in Eq. (\ref{eaasy26}). Due to the large cancellation among different
terms in $z$--series expansion, we could not fix all coefficients $\beta_k$
from the fit. Therefore, we keep only the first two
terms in the expansion, i.e., we take $N=2$ (see also \cite{Raasy19}).
    
\begin{table}[tbh]

\renewcommand{\arraystretch}{1.3}
\addtolength{\arraycolsep}{-0.5pt}
\small
$$
\begin{array}{|l|c|c|c|}
\hline \hline
     & \beta_{0}^{K_{1A}} & \beta_{1}^{K_{1A}}  & \sum_{k=0}^1{(\beta_k^{K_{1A}})^2} \\  \hline \hline
 H_0 & 3.7\times 10^{-5}  & -1.3\times 10^{-3}  & 1.7\times 10^{-6}       \\
 H_1 & 8.4\times 10^{-5}  & -3.0\times 10^{-3}  & 9.0\times 10^{-6}       \\
 H_2 & 2.1\times 10^{-5}  & -6.4\times 10^{-4}  & 4.1\times 10^{-7}       \\
\hline \hline
\end{array}
$$
\caption{The values of the coefficients $\beta_k^{K_{1A}}$ of 
the series expansion parametrization of the  helicity amplitudes 
$H_0(q^2)$, $H_1(q^2)$ and $H_2(q^2)$.}
\renewcommand{\arraystretch}{1}
\addtolength{\arraycolsep}{-1.0pt}
\end{table}

Our final remark is as follows: As has already been noted, in the numerical
calculations we use $f_B=210 \pm 19~MeV$. If $f_B=145~MeV$ \cite{Raasy20} had been used the
values of all form factors presented in this work increase by a factor 1.4.

In summary, we calculate the tensor form factors of $B$
decays into $P$--wave axial--vector meson. These form factors are relevant
to the studies for the exclusive FCNC transitions. The sum rules obtained
could further be improved by including the ${\cal O}(\alpha_s)$ corrections,
as well as, improving the values of the input parameters involving the DAs.

\newpage

\section*{Acknowledgments}

%\begin{\acknowledgments}
This research of K. C. Y. is supported in part by the National Center for Theoretical Sciences and the
National Science Council of R.O.C. under Grant No. NSC99-2112-M-003-005-MY3.
%\end{acknowledgments}

\newpage

\appendix
\renewcommand{\theequation}{A.\arabic{equation}}
\section*{Appendix}
\section*{Distribution amplitudes}
\setcounter{equation}{0}

The two--parton chiral--even LCDAs are given by
\begin{eqnarray}
  \langle K_1(p,\lambda)|\bar s(x) \gamma_\mu \gamma_5 \psi(0)|0\rangle
  & =& if_{K_1} m_{K_1} \, \int_0^1
      du \,  e^{i u \, p x}
   \Bigg\{ p_\mu \,
    \frac{\varepsilon^{(\lambda)*} x}{p x} \, \phi_\parallel(u) 
\nonumber\\
 & & + \left( \varepsilon_{\mu}^{(\lambda)*} -p_\mu \frac{\varepsilon^{(\lambda)*} x}{p\,x}\right)\,
    g_\perp^{(a)}(u)
  - \frac{1}{2}x_{\mu} \frac{\epsilon^{*(\lambda)} x }{(p  x)^{2}} m_{K_1}^{2} \bar g_{3}(u) 
 \nonumber\\
 & &  + {\cal O}(x^2) \Bigg\}~,
                                 \label{eq:evendef1} \\
  \langle K_1 (p,\lambda)|\bar s(x) \gamma_\mu \psi(0)|0\rangle
  & = & - i f_{K_1} m_{K_1}
\,\epsilon_{\mu\nu\rho\sigma} \,
      \varepsilon^{(\lambda)*\nu} p^{\rho} x^\sigma \, \int_0^1 du \,  e^{i u \, p x} \,
      \Bigg( \frac{g_\perp^{(v)}(u)}{4} + {\cal O}(x^2)\Bigg)~,\nonumber\\
       \label{eq:evendef2}
\end{eqnarray}
where $\psi\equiv u ({\rm or}\ d)$, $x^2\not= 0$, and  $u$ is the momentum fraction carried by the
$s$ in the $K_{1A(B)}$ meson. The two--parton chiral--odd LCDAs are defined by
\begin{eqnarray}
  &&\langle K_1(p,\lambda)|\bar s(x) \sigma_{\mu\nu}\gamma_5 \psi(0)
            |0\rangle
  =  f_{K_1}^{\perp} \,\int_0^1 du \, e^{i u \, p x} \,
\Bigg\{(\varepsilon^{(\lambda)*}_{\mu} p_{\nu} -
  \varepsilon_{\nu}^{(\lambda)*}  p_{\mu})
  \phi_\perp(u)\nonumber\\
&& \hspace*{+5cm}
  + \,\frac{m_{K_1}^2\,\varepsilon^{(\lambda)*} x}{(p x)^2} \,
   (p_\mu x_\nu - p_\nu  x_\mu) \, \bar{h}_\parallel^{(t)}(u)\nonumber\\
 && \hspace*{+5cm} + \frac{1}{2}
(\varepsilon^{*(\lambda)}_{\mu} x_\nu
-\varepsilon^{*(\lambda)}_{\nu} x_\mu) \frac{m_{K_1}^{2}}{p x}  \bar{h}_{3}(u)
 +{\cal O}(x^2)
\Bigg\}~,\label{eq:odddef1}\\
&&\langle K_1(p,\lambda)|\bar s(x) \gamma_5 \psi(0) |0\rangle
  =  f_{K_1}^\perp
 m_{K_1}^2 (\varepsilon^{*(\lambda)} x)\,\int_0^1 du \, e^{i u \, p x}  \,
\Bigg(\frac{h_\parallel^{(p)}(u)}{2}+ {\cal O}(x^2)\Bigg)~,
\label{eq:odddef2}
\end{eqnarray}
where the functions $\bar{g}_3 (u)$, $\bar{h}_\parallel^{(t)}$ and
$\bar{h}_3 (u)$ are given in Eq. (\ref{eaasy19}).
In SU(3) limit, due to $G$-parity,
$\phi_\parallel, g_\perp^{(a)}$, $g_\perp^{(v)}$, and $g_3$ are
symmetric (antisymmetric) under the replacement $u\to 1-u$ for the
$1^3P_1$ ($1^1P_1$) states, whereas $\phi_\perp, h_\parallel^{(t)}$,
$h_\parallel^{(p)}$, and $h_3$ are antisymmetric (symmetric). Up to twist--3, we adopt the following normalization conventions \cite{Raasy14},
\begin{eqnarray}
 & & \int_0^1 du \phi_\parallel(u)=\int_0^1 du g_\perp^{(a)}(u) =\int_0^1 du g_\perp^{(v)}(u)=1, \nonumber\\
 & & \int_0^1 du \phi_\perp(u)=\int_0^1 du h_\parallel^{(t)}(u) =a_0^\perp, \quad
 \int_0^1 du h_\parallel^{(p)}(u) = a_0^\perp+\delta_-~,
 \end{eqnarray}
for $K_{1A}$, but becomes
\begin{eqnarray}
 & & \int_0^1 du \phi_\parallel(u)=\int_0^1 du g_\perp^{(a)}(u) =a_0^\parallel, \quad
      \int_0^1 du g_\perp^{(v)}(u)=a_0^\parallel + \tilde\delta_-~, \nonumber\\
 & & \int_0^1 du \phi_\perp(u)=\int_0^1 du h_\parallel^{(t)}(u) = \int_0^1 du h_\parallel^{(p)}(u) =1,
 \end{eqnarray}
for $K_{1B}$, where 
\begin{equation}
\widetilde{\delta}_-  =- {f_{K_1}\over f_{K_1}^{\perp}}{m_{s}
\over m_{K_1}}~,
\qquad 
  \widetilde{\delta}_-  =- {f_{K_1}^{\perp}\over f_{K_1}}{m_{s} \over m_{K_1}}~,
\label{eq:parameters3}
\end{equation}
and $a_0^{\parallel,\perp}$ are defined through
 \begin{eqnarray}
 &&\langle K_{1A}(p,\lambda)|
  \bar s(0) \sigma_{\mu\nu}\gamma_5 \psi(0) |0\rangle
  =  f_{K_{1A}} a_0^{\perp,K_{1A}} \,
(\varepsilon^{(\lambda)*}_{\mu} p_{\nu} - \varepsilon_{\nu}^{(\lambda)*} p_{\mu})~,
 \\
 &&  \langle K_{1B}(p,\lambda)|\bar s(0) \gamma_\mu \gamma_5 \psi(0)|0\rangle
   = if_{K_{1B}}^{\perp}(1~{\rm GeV}) a_0^{\parallel, K_{1B}} \, m_{K_{1B}} \,
    \varepsilon^{(\lambda)*}_\mu
   ~,
 \end{eqnarray}
with $f_{K_{1A}}^\perp =f_{K_{1A}}$ and $f_{^1\! P_1}=f_{^1\!
P_1}^\perp(\mu=1~{\rm GeV})$. $a_0^{\perp,K_{1A}}$ and  $a_0^{\parallel,K_{1B}}$ are the
G--parity violating
zeroth Gegenbauer moments and vanish in the SU(3) limit. 

We use the twist--2 distributions
\cite{Raasy14}
\begin{eqnarray}
\phi_\parallel(u) & = & 6 u \bar u \left[ 1 + 3 a_1^\parallel\, \xi +
a_2^\parallel\, \frac{3}{2} ( 5\xi^2  - 1 )
 \right]~, \label{eq:lcda-3p1-t2-1}\\
 \phi_\perp(u) & = & 6 u \bar u \left[ a_0^\perp + 3 a_1^\perp\, \xi +
a_2^\perp\, \frac{3}{2} ( 5\xi^2  - 1 ) \right]~, \label{eq:lcda-3p1-t2-2}
\end{eqnarray}
for the $K_{1A}$ and 
\begin{eqnarray}
 \phi_\parallel(u) & = & 6 u \bar u \left[ a_0^\parallel + 3
a_1^\parallel\, \xi +
a_2^\parallel\, \frac{3}{2} ( 5\xi^2  - 1 ) \right]~, \label{eq:lcda-1p1-t2-1}\\
\phi_\perp(u) & = & 6 u \bar u \left[ 1 + 3 a_1^\perp\, \xi +
a_2^\perp\, \frac{3}{2} ( 5\xi^2  - 1 ) \right]~,
\label{eq:lcda-1p1-t2-2}
\end{eqnarray}
for the $K_{1B}$, where $\xi=2u-1$. For the relevant three-parton twist--3
chiral--even LCDAs, we use
the contributions up to terms of conformal spin $9/2$ and take into account the corrections arising from the strange quark mass:
 \begin{eqnarray}
  {\cal A}&=&5040 (\alpha_1-\alpha_2)\alpha_1\alpha_2\alpha_3^2
  +360\alpha_1\alpha_2\alpha_3^2
  \Big[ \lambda^A_{K_{1A}}+ \sigma^A_{K_{1A}}\frac{1}{2}(7\alpha_3-3)\Big]~,
 \label{eq:lcda-3p1-t3-1}\\
  {\cal V}&=&360\alpha_1\alpha_2\alpha_3^2
  \Big[ 1+ \omega^V_{K_{1A}}\frac{1}{2}(7\alpha_3-3)\Big]
 +5040 (\alpha_1-\alpha_2)\alpha_1\alpha_2\alpha_3^2
 \sigma^V_{K_{1A}}~,\label{eq:lcda-3p1-t3-2}
 \end{eqnarray}
for the $K_{1A}$, and
 \begin{eqnarray}
  {\cal A}&=&360 \alpha_1\alpha_2\alpha_3^2
  \Big[ 1+ \omega^A_{K_{1B}}\frac{1}{2}(7\alpha_3-3)\Big]
  +5040 (\alpha_1-\alpha_2)\alpha_1\alpha_2\alpha_3^2
 \sigma^A_{K_{1B}}~, \label{eq:lcda-1p1-t3-1}\\
  {\cal V}&=&5040 (\alpha_1-\alpha_2)\alpha_1\alpha_2\alpha_3^2
  +360\alpha_1\alpha_2\alpha_3^2
  \Big[ \lambda^V_{K_{1B}}+ \sigma^V_{K_{1B}}\frac{1}{2}(7\alpha_3-3)\Big]~,
  \label{eq:lcda-1p1-t3-2}
 \end{eqnarray}
for the $K_{1B}$, where $\lambda$'s correspond to conformal
spin 7/2, while $\omega$'s and $\sigma$'s are parameters with
conformal spin 9/2. As the SU(3)--symmetry is
restored, we have $\lambda$'s=$\sigma$'s=0.

 For the relevant two--parton twist--3 chiral--even LCDAs, we take the approximate
expressions up to conformal spin $9/2$ and of order $m_s$ \cite{Raasy14}:
\begin{eqnarray}
 g_\perp^{(a)}(u) & = &  \frac{3}{4}(1+\xi^2)
+ \frac{3}{2}\, a_1^\parallel\, \xi^3
 + \left(\frac{3}{7} \,
a_2^\parallel + 5 \zeta_{3,K_{1A}}^V \right) \left(3\xi^2-1\right)
 \nonumber\\
& & {}+ \left( \frac{9}{112}\, a_2^\parallel + \frac{105}{16}\,
 \zeta_{3,K_{1A}}^A - \frac{15}{64}\, \zeta_{3,K_{1A}}^V \omega_{K_{1A}}^V
 \right) \left( 35\xi^4 - 30 \xi^2 + 3\right) \nonumber\\
 & &
 + 5\Bigg[ \frac{21}{4}\zeta_{3,K_{1A}}^V \sigma_{K_{1A}}^V
  + \zeta_{3,K_{1A}}^A \bigg(\lambda_{K_{1A}}^A -\frac{3}{16}
 \sigma_{K_{1A}}^A\Bigg) \Bigg]\xi(5\xi^2-3)
 \nonumber\\
& & {}-\frac{9}{2} {a}_1^\perp
\,\widetilde{\delta}_+\,\left(\frac{3}{2}+\frac{3}{2}\xi^2+\ln u
 +\ln\bar{u}\right) - \frac{9}{2} {a}_1^\perp\,\widetilde{\delta}_-\, (
3\xi + \ln\bar{u} - \ln u)~, \label{eq:ga-3p1}\\
g_\perp^{(v)}(u) & = & 6 u \bar u \Bigg\{ 1 +
 \Bigg(a_1^\parallel + \frac{20}{3} \zeta_{3,K_{1A}}^A
 \lambda_{K_{1A}}^A\Bigg) \xi\nonumber\\
 && + \Bigg[\frac{1}{4}a_2^\parallel + \frac{5}{3}\,
 \zeta^V_{3,K_{1A}} \left(1-\frac{3}{16}\, \omega^V_{K_{1A}}\right)
 +\frac{35}{4} \zeta^A_{3,K_{1A}}\Bigg] (5\xi^2-1) \nonumber\\
 &&+ \frac{35}{4}\Bigg(\zeta_{3,K_{1A}}^V
 \sigma_{K_{1A}}^V -\frac{1}{28}\zeta_{3,K_{1A}}^A
 \sigma_{K_{1A}}^A \Bigg) \xi(7\xi^2-3) \Bigg\}\nonumber\\
& & {} -18 \, a_1^\perp\widetilde{\delta}_+ \,  (3u \bar{u} +
\bar{u} \ln \bar{u} + u \ln u ) - 18\,
a_1^\perp\widetilde{\delta}_- \,  (u \bar u\xi + \bar{u} \ln \bar{u} -
u \ln u)~,
 \label{eq:gv-3p1}
 \end{eqnarray}
for the $K_{1A}$, and
\begin{eqnarray}
 g_\perp^{(a)}(u) & = & \frac{3}{4} a_0^\parallel (1+\xi^2)
+ \frac{3}{2}\, a_1^\parallel\, \xi^3
 + 5\left[\frac{21}{4} \,\zeta_{3,K_{1B}}^V
 + \zeta_{3,K_{1B}}^A \Bigg(1-\frac{3}{16}\omega_{K_{1B}}^A\Bigg)\right]
 \xi\left(5\xi^2-3\right)
 \nonumber\\
& & {}+ \frac{3}{16}\, a_2^\parallel \left(15\xi^4 -6 \xi^2 -1\right)
 + 5\, \zeta^V_{3,K_{1B}}\lambda^V_{K_{1B}}\left(3\xi^2 -1\right)
 \nonumber\\
& & {}+ \frac{105}{16}\left(\zeta^A_{3,K_{1B}}\sigma^A_{K_{1B}}
-\frac{1}{28} \zeta^V_{K_{1B}}\sigma^V_{K_{1B}}\right)
 \left(35\xi^4 -30 \xi^2 +3\right)\nonumber\\
 & & {}-15 {a}_2^\perp \bigg[ \widetilde{\delta}_+ \xi^3 +
 \frac{1}{2}\widetilde{\delta}_-(3\xi^2-1) \bigg] \nonumber\\
& & {}
  -\frac{3}{2}\,\bigg[\widetilde{\delta}_+\, ( 2 \xi + \ln\bar{u} -\ln u)
 +\, \widetilde{\delta}_-\,(2+\ln u + \ln\bar{u})\bigg](1+6a_2^\perp)
 ~,\label{eq:ga-1p1}\\
g_\perp^{(v)}(u) & = & 6 u \bar u \Bigg\{ a_0^\parallel +
a_1^\parallel \xi +
 \Bigg[\frac{1}{4}a_2^\parallel
  +\frac{5}{3} \zeta^V_{3,K_{1B}}
  \Bigg(\lambda^V_{K_{1B}} -\frac{3}{16} \sigma^V_{K_{1B}}\Bigg)
  +\frac{35}{4} \zeta^A_{3,K_{1B}}\sigma^A_{K_{1B}}\Bigg](5\xi^2-1) \nonumber\\
  & & {}  + \frac{20}{3}\,  \xi
 \left[\zeta^A_{3, K_{1B}}
 + \frac{21}{16}
 \Bigg(\zeta^V_{3,K_{1B}}- \frac{1}{28}\, \zeta^A_{3,K_{1B}}\omega^A_{K_{1B}}
  \Bigg)
 (7\xi^2-3)\right]\nonumber\\
 & & {} -5\, a_2^\perp [2\widetilde\delta_+ \xi + \widetilde\delta_- (1+\xi^2)]
 \Bigg\}\nonumber\\
 & & {} - 6 \bigg[\, \widetilde{\delta}_+ \, (\bar{u} \ln\bar{u} -u\ln u )
  +\, \widetilde{\delta}_- \, (2u \bar{u} + \bar{u} \ln \bar{u} + u \ln u)\bigg]
  (1+6 a_2^\perp)~,
 \label{eq:gv-1p1}
\end{eqnarray}
for the $K_{1B}$, where
\begin{equation}
\widetilde{\delta}_\pm  =\pm {f_{K_1}^{\perp}\over f_{K_1}}{m_{s}
\over m_{K_1}}~,\qquad \zeta_{3, K_1}^{V(A)} = \frac{f^{V(A)}_{3 K_1}}{f_{K_1}
m_{K_1}}~.
\label{eq:parameters3}
\end{equation}

The relevant parameters entering to the expressions of DAs are listed in
Table 3.

%%%%%%%%%%%%%%%%%%%%%%%%%%%%%%%%%%%%%%%%%%%%%%%%%%%%%%%%%%%%%%%%%%%%%%%%%%%%%%%
%
\begin{table}[ht!]
\centerline{\parbox{16cm}{\caption{\label{tab:inputs} Gegenbauer moments of
twist--2 and twist--3 LCDAs at the scale 2.2 GeV \cite{Raasy14}. The G--parity violating 
parameters are updated using new values for $a_0^{\perp, K_{1A}}$ and 
$a_0^{\parallel, K_{1B}}$ given in Ref. \cite{Raasy15}.
}}}
\begin{center}
{\tabcolsep=0.194cm\begin{tabular} {|c| c c c c c c|}\hline\hline
\multicolumn{7}{|c|}{Gegenbauer moments for leading--twist LCDAs}
 \\
 \hline  & $a_0^\parallel$  & $a_1^\parallel$ & $a_2^\parallel$ & $a_0^\perp$ & $a_1^\perp$ & $a_2^\perp$\\
 $K_{1A}$ & $1$ & $-0.25^{+0.00}_{-0.17}$ & $-0.04\pm0.02$ & $0.25^{+0.03}_{-0.16}$ & $-0.88\pm0.39$ 
 & $0.01\pm0.15$\\
  $K_{1B}$ & $-0.19\pm0.07$ & $-1.57\pm0.37$ & $0.07^{+0.11}_{-0.14}$ & $1$ & $0.24^{+0.00}_{-0.27}$
 & $-0.02\pm0.17$\\
\hline
\end{tabular}}
{\small{\tabcolsep=0.585cm\begin{tabular}{|c| c c c|} \hline
\multicolumn{4}{|c|}{G-parity conserving parameters of twist--3 3--parton LCDAs}
 \\
\hline
 &~~~~~~~$f^V_{3,^3P_1}$ [GeV$^2$]~~~~~& ~~~~~~~~~~ $\omega_{^3P_1}^V$~~~~~~~~
  & $f^A_{3,^3P_1}$ [GeV$^2$] \\
  ~~~~~$K_{1A}$~~~~ & $0.0034\pm 0.0018$&$-3.1  \pm 1.1$ & $0.0014\pm 0.0007$ \\
  \hline
  &~~~~~~~$f^A_{3,^1P_1}$ [GeV$^2$]~~~~~& ~~~~~~~~~~ $\omega_{^1P_1}^A$~~~~~~~~~~
  & $f^V_{3,^1P_1}$ [GeV$^2$] \\
  ~~~~~$K_{1B}$~~~~& $-0.0041\pm 0.0018$& $-1.7  \pm 0.4$ & $0.0029\pm 0.0012$ \\
  \hline
\end{tabular}}
}
{\small{\tabcolsep=1.134cm\begin{tabular}{|ccc|} \hline
\multicolumn{3}{|c|}{G-parity violating parameters of twist--3 3--parton LCDAs of the $K_{1A}$}
 \\
\hline
 ~~~~~~~~~~~~$\sigma_{K_{1A}}^V$~~~~~~~~~~~~~
&~~~~~~~~~ $\lambda_{K_{1A}}^A$~~~~~~~ & ~~~~~~$\sigma_{K_{1A}}^A$~~~~~~ \\
 $0.01\pm 0.04$ &  $-0.12  \pm 0.22$ & $-1.9\pm 1.1$\\
\hline
\end{tabular}}
}
{\small{\tabcolsep=1.086cm\begin{tabular}{|ccc|} \hline
\multicolumn{3}{|c|}{G-parity violating parameters of twist--3 3--parton LCDAs of the $K_{1B}$}
 \\
\hline
 ~~~~~~~~~~~~ $\lambda_{K_{1B}}^V$~~~~~~~~~~~
&~~~~~~~~~~~$\sigma_{K_{1B}}^V$~~~~~~~~ & ~~~~~~~$\sigma_{K_{1B}}^A$~~~~~~ \\
 $-0.23 \pm 0.18$ & $ 1.3\pm 0.8$ &  $0.03\pm 0.03$\\
\hline\hline
\end{tabular}}}
\vskip0.8cm
\end{center}
\end{table}

\newpage
~~~~~~\\ \\ \\ \\  
\begin{figure}[tbh]
\vskip 1.5 cm
    \includegraphics{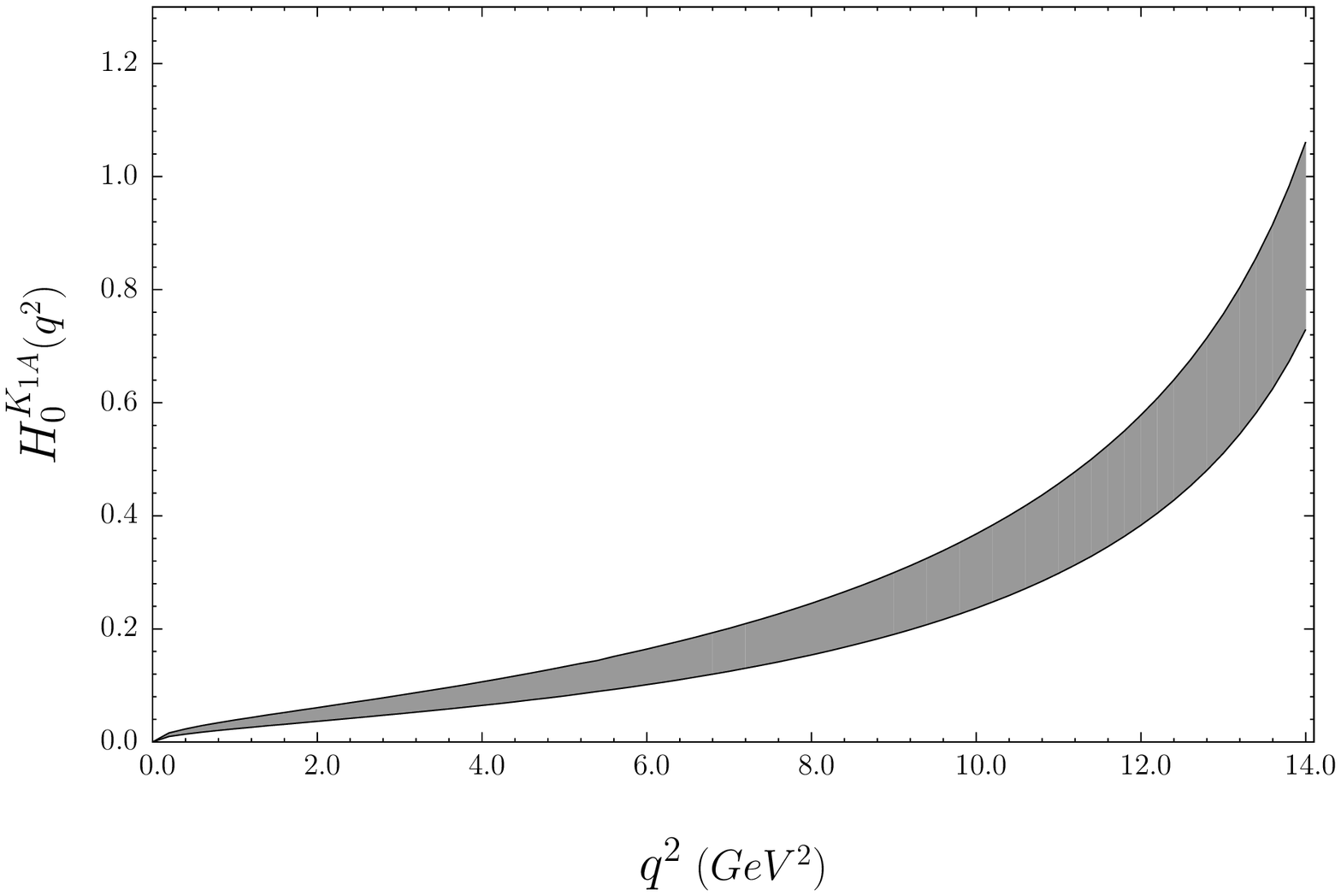}
\vskip 8.5cm
\caption{The dependence of the helicity amplitude $H_0(q^2)$ on $q^2$ for
the $z$--series expansion parametrization fitted to the LCSR prediction for
the $B \to K_{1A}$ transition}
%\begin{center}
%{\bf Fig. 1--a}
%\end{center}
\end{figure}

\begin{figure}[tbh]
\vskip 1.5 cm
    \includegraphics{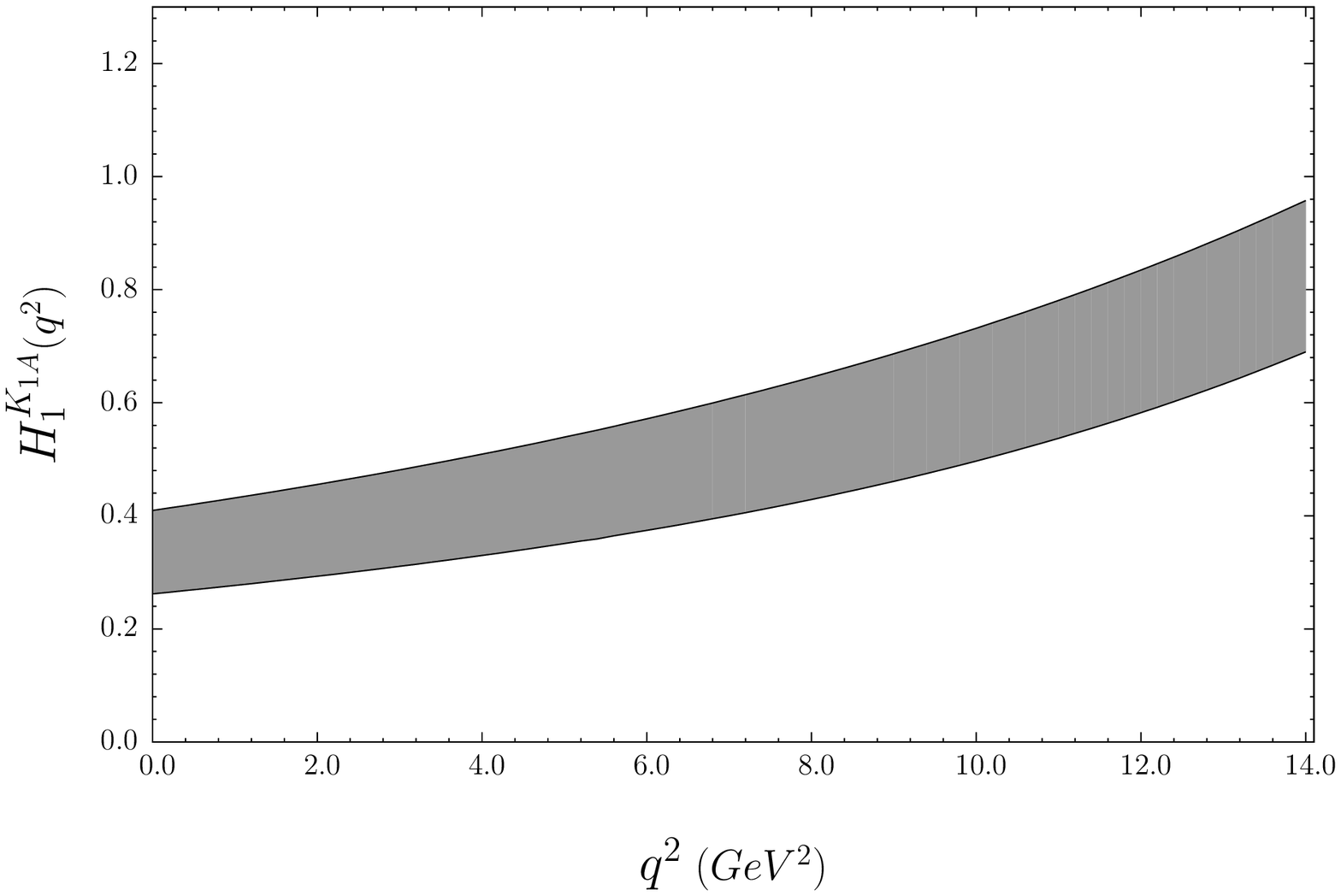}
\vskip 8.5cm
\caption{The same as in Fig. (1), but for the helicity amplitude $H_1(q^2)$.}
%\begin{center}
%{\bf Fig. 1--a}
%\end{center}
\end{figure}

\begin{figure}[tbh]
\vskip 1.5 cm
    \includegraphics{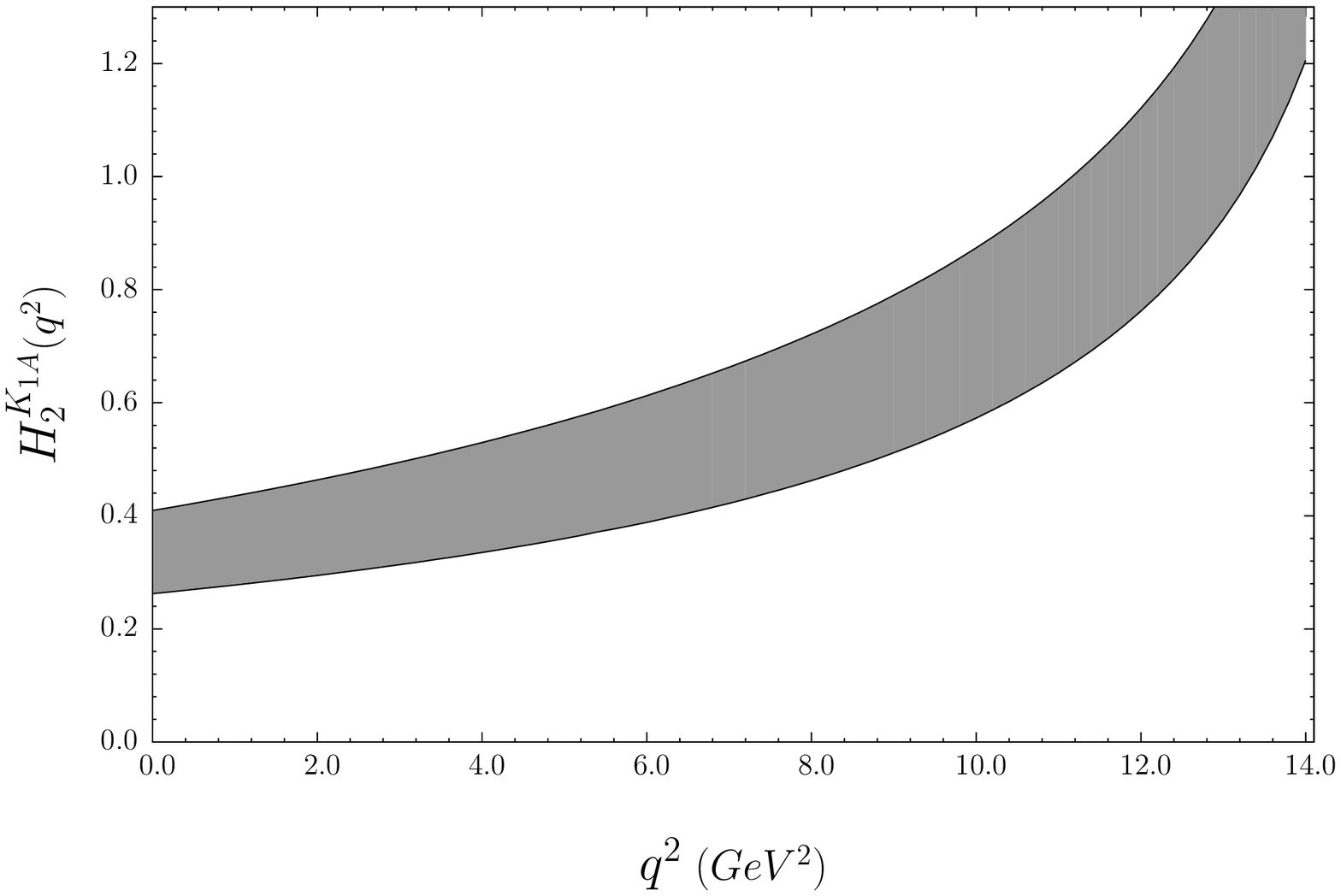}
\vskip 8.5cm
\caption{The same as in Fig. (1), but for the helicity amplitude $H_2(q^2)$.}
%\begin{center}
%{\bf Fig. 1--a}
%\end{center}
\end{figure}

\end{document}